\documentclass[useAMS,letters,usegraphicx,usenatbib]{mn2e}

\title[VLBI observation of the newly discovered $z=5.18$ quasar]{VLBI observation of the newly discovered $z=5.18$ quasar SDSS J0131$-$0321}
\author[K. \'E. Gab\'anyi et al.]{K. \'E. Gab\'anyi$^{1,2}$\thanks{E-mail:
gabanyik@sgo.fomi.hu; krisztina.g@gmail.com}, D. Cseh$^{3}$, S. Frey$^{1}$, Z. Paragi$^{4}$, L. I. Gurvits$^{4,5}$, T. An$^{6,7}$and \newauthor Y. K. Zhang$^{6,7}$\\
$^{1}$F\"OMI Satellite Geodetic Observatory, P.O. Box 585, 1592 Budapest, Hungary\\
$^{2}$Konkoly Observatory, MTA Research Centre for Astronomy and Earth Sciences, P.O. Box 67, 1525 Budapest, Hungary\\
$^{3}$Deptartment of Astrophysics/IMAPP, Radboud University Nijmegen, P.O. Box 9010, 6500 GL Nijmegen, The Netherlands\\
$^{4}$Joint Institute for VLBI in Europe, P.O. Box 2, 7990 AA Dwingeloo, The Netherlands\\ 
$^{5}$Deptartment of Astrodynamics and Space Missions, Delft University of Technology, Kluyverweg 1, 2629 HS Delft, The Netherlands\\
$^{6}$Shanghai Astronomical Observatory, Chinese Academy of Sciences, 80 Nandan Road, 200030 Shanghai, P. R. China\\
$^{7}$Key Laboratory of Radio Astronomy, Chinese Academy of Sciences, 210008 Nanjing, P. R. China\\
}
\begin{document} 

\date{Accepted 2015 . Received 2015 ; in original form 2015}

\pagerange{\pageref{firstpage}--\pageref{lastpage}} \pubyear{2015}

\maketitle

\label{firstpage}

\begin{abstract}
Few high-redshift, radio-loud quasars are known to date. The extremely luminous, radio-bright quasar, SDSS J013127.34$-$032100.1 was recently discovered at a redshift of $z=5.18$. We observed the source with high resolution very long baseline interferometry (VLBI) at 1.7\,GHz with the European VLBI Network (EVN) and found a single compact radio component. We estimated a lower limit to the brightness temperature of the detected radio component, $T_\rmn{B}\sim10^{11}$\,K. Additionaly, when compared to archival radio data, the source showed significant flux density variation. These two findings are indicative of the blazar nature of the source.
\end{abstract}


\section{Introduction}
High-redshift ($z\ga4.5$) quasars are of major importance since they can provide information on the growth of the supermassive black holes and the evolution of active galactic nuclei (AGN) in the early Universe. The mere existence of black holes with a few million solar masses (or more) at $z\sim6$ constrains the black hole growth and the accretion process \citep[e.g.][and references therein]{volonteri}. Among the high-redshift quasars, the radio-loud ones constitute a very attractive subsample, since their radio jets can be studied with the highest angular resolution, via very long baseline interferometry (VLBI) technique. 

Typically, radio-loud AGN contain a flat-spectrum core and a steep-spectrum jet. However, the higher redshift is, the fainter the AGN radio jet will appear if observed at a fixed wavelength. Thus, if core-jet AGN constitute the same population of objects throughout the redshift space, the apparent ``prominence'' of jets at higher redshifts must decrease \citep{gurvits}: well pronounced jets at high redshifts must appear less frequent than at low redshifts. Additionaly, recent VLBI studies suggest that the naive expectation that relativistically beamed sources (blazars) should dominate the high-redshift radio-loud quasar population might not be true. There seems to exist a population of steep-spectrum high-redshift ($z\ga4.5$) radio-loud quasars \citep{double_Momjian,double_Frey,fiveqso}, which may be very young radio quasars, similar to the gigahertz-peaked spectrum (GPS) sources observed in the local Universe \citep{fiveqso}.
On the other hand, a few high-redshift, beamed blazars are known as well.
Three are identified at $z\ga5$: SDSS J114657.79+403708.6 \citep{fiveqso, 1146_Ghisellini}, SDSS J102623.61+254259.5 \citep{1026_Sbarrato,1026_Frey}, and Q0906+6930 \citep{0906_Romani,cand_Sbarrato}.

Recently \citet{disc} reported the discovery of a high-redshift, radio-bright quasar, SDSS J013127.34$-$032100.1 (J0131$-$0321, hereafter). The source was first selected as a candidate high-redshift quasar \citep{sdss_selection} using the optical--infrared selection criteria based on the Sloan Digital Sky Survey \citep[SDSS,][]{sdss} and {\it Wide-Field Infrared Survey Explorer} \citep[{\it WISE},][]{wise} photometry. Its optical spectrum was first measured with the Yunnan Fainter Object Spectrograph and Camera \citep{FOSC} 
in 2013 November. Later higher resolution optical and near-infrared spectra were obtained by the Magellan Echelette and Folded Port Infrared Echelette (FIRE) spectrographs in 2014 January. According to these measurements, the source has a redshift of $z=5.18 \pm 0.01$ \citep{disc}.

Using the optical spectra and the empirical scaling relation between the Mg {\sc II} line width and the black hole mass \citep{mclure}, \citet{disc} estimated the total bolometric luminosity and the black hole mass of J0131$-$0321 to be $L_\rmn{bol}=(1.1\pm0.2)\times10^{41}$\,W and $M_\rmn{BH}=2.7^{+0.5}_{-0.4}\times10^9 M_\odot$, respectively. Using the latest relation of \citet{Trak}, the black hole mass can be even higher, $4\times10^9 M_\odot$. 

According to the Faint Images of the Radio Sky at Twenty Centimeters \citep[FIRST,][]{first} survey, J0131$-$0321 is radio-bright with a flux density of 33.7\,mJy at 1.4\,GHz. Here we report on the results of our high resolution radio interferometric observation of the source with the European VLBI Network (EVN).

Throughout this paper, we use flat $\Lambda$ cold dark matter cosmological model with $H_0=70$\,km\,s$^{-1}$\,Mpc$^{-1}$, $\Omega_\rmn{m}=0.3$, $\Omega_\Lambda=0.7$ \citep[the same as used in the discovery paper,][]{disc}. 

\section{VLBI Observation and Data Reduction}
The exploratory EVN observation of J0131$-$0321 took place on 2014 Dec 2 at 1.7\,GHz (project code: RSG06). Six antennas participated in this e-VLBI experiment: Effelsberg (Germany), Hartebeesthoek (South Africa), Jodrell Bank Mk II (United Kingdom), Onsala (Sweden), Toru\'n (Poland), Sheshan (China). In an e-VLBI experiment \citep{eEVN}, the signals received at the remote radio telescopes are transmitted over optical fiber networks directly to the central data processor for real-time correlation. This was performed at the EVN software correlator \citep[SFXC,][]{soft_corr} in the Joint Institute for VLBI in Europe (JIVE), Dwingeloo, the Netherlands, with 2 s integration time. The observation was carried out in phase-reference mode \citep{phase-ref}, thus precise relative positional information could be obtained. The phase-reference calibrator, J0123$-$0348 is separated by $\sim 2^\circ$ from the target in the sky. Its coordinates are right ascension $\alpha_0=01^\rmn{h}23^\rmn{m}35\fs 77473$ and declination $\delta_0=-03^\circ48\arcmin 39\farcs 3162$. The positional uncertainties are $0.3$ and $0.63$ milliarcseconds (mas) in right ascension and declination, respectively\footnote{Data are from http://astrogeo.org maintained by L. Petrov, rfc\_2014d solution}.  The observation lasted for 2 h, the total time spent on J0131$-$0321 was nearly $90$\,min. Eight intermediate frequency channels (IFs) were used in both right and left circular polarizations. Each IF was divided into 64 spectral channels. The total bandwidth was $128$\,MHz.

The NRAO Astronomical Image Processing System \citep[{\sc aips},][]{aips} was used for data calibration \citep{data_reduc} following standard procedures. The interferometric visibility amplitudes were calibrated using the system temperatures and antenna gain curves measured at the telescopes. After fringe-fitting performed for the phase calibrator, its visibility data were exported to {\sc difmap} \citep{difmap} for imaging. The standard hybrid mapping procedure was used, with several cycles of {\sc clean}ing \citep{clean} and phase self-calibration, then finally amplitude self-calibration. Antenna gain correction factors were determined in {\sc difmap} and applied to the data in {\sc aips}. The correction factors on average were below $10$ per cent. Fringe-fitting was repeated for the calibrator source but now taking into account its {\sc clean} component model (representative of its brightness distribution) to reduce the small phase variations due to its structure. The obtained solutions were interpolated and applied to the target source, J0131$-$0321. In order to correct for the bandpass, four-four channels were discarded at the edges of each IF. Then the calibrated and phase-referenced data of the target source were imaged in {\sc difmap}. No amplitude self-calibration was attempted for the target source.
Since J0131$-$0321 turned out to be adequately bright (see Sect. \ref{result}), we were able to fringe-fit the source itself in {\sc aips}. The images obtained with and without using the phase-calibrator source agree within the uncertainties. In Fig. \ref{fig:EVN}, we show the image made from the fringe-fitted visibility data. The weights of the data points were set inversely proportional to the amplitude errors (natural weighting).

\section{Results of the EVN observation}
\label{result}

\begin{figure}
\includegraphics[angle=-90,width=\columnwidth]{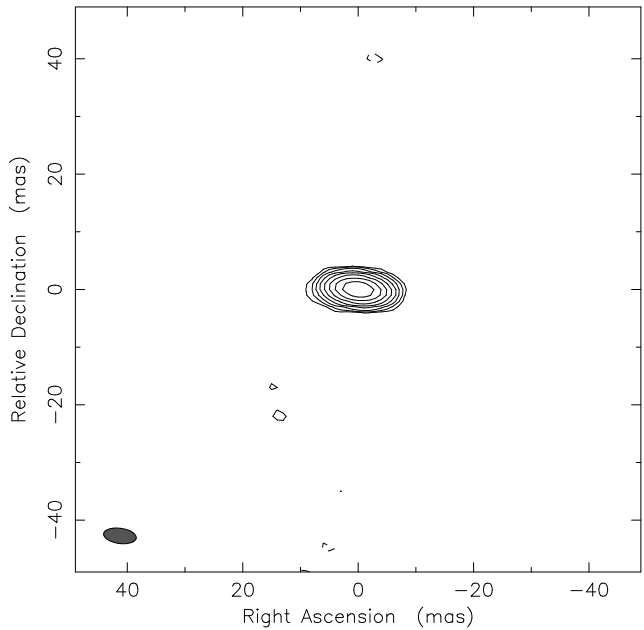}
\caption{1.7-GHz EVN image of J0131$-$0321. The peak is $60.1$\,mJy\,beam$^{-1}$, the restoring beam is $5.7\rmn{\,mas}\times 2.7$\,mas at a position angle of $82^\circ$ and shown in the lower left corner of the image. The lowest contours are at $\pm3\,\sigma$ noise level ($\pm 0.25$\,mJy\,beam$^{-1}$), further positive contour levels increase by a factor of two.}
\label{fig:EVN}
\end{figure}

\begin{figure}
\includegraphics[width=70mm,angle=-90]{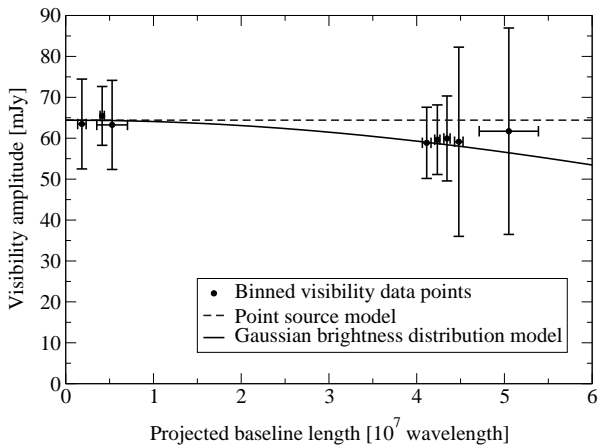}
\caption{Calibrated visibility amplitudes of J0131$-$0321 as a function of the projected interferometer baseline length. The visibility points are binned for illustration purposes only, model fitting was performed by using all visibility points. The binned amplitudes are weighted averages of the visibility points (see text for details), the horizontal error bars represent the width of each bin, the vertical error bars are the square root of the weighted sample variance. Solid line shows the best fit circular Gaussian model, dashed line shows the visibility amplitude of an unresolved point source with the same flux density as the Gaussian component.}
\label{fig:radplot}
\end{figure}

Our EVN observation revealed a single, compact radio component (Fig. \ref{fig:EVN}). Using the phase-referenced data, we derived the position of the brightness peak with the {\sc maxfit} verb in {\sc aips}: right ascension $\alpha_\rmn{t}=01^\rmn{h}31^\rmn{m}27\fs3473$ and declination $\delta_\rmn{t}=-03^\circ21\arcmin 00\farcs 0791$. We estimate that each coordinate is accurate within 1\,mas. The dominant contribution to the error is the positional uncertainty of the phase-reference calibrator. Additional errors come from the thermal noise of the interferometer phases, and from the phase-referencing over $2^\circ$ in the sky. The obtained position is $\sim 117$\,mas away from the coordinates given in the FIRST catalogue. 

To quantify the source brightness distribution, we fitted the self-calibrated visibilities with a single circular Gaussian model component in {\sc difmap}. The integrated flux density is $(64.4\pm 0.3)$\,mJy, the full width at half maximum (FWHM) size is $(0.79 \pm 0.01)$\,mas. The statistical errors were calculated following the formulae of \citet{error}. The angular size corresponds to a projected linear size of $\sim4.9$\,pc in the rest frame of the source. The fitted size of the Gaussian component is close to, but slightly larger than the minimum resolvable size \citep[see][]{kovalev}. 

The calibrated visibility amplitudes as the function of the projected baseline length are shown in Fig. \ref{fig:radplot} together with the fitted Gaussian brightness distribution model (solid line) and a point source model with the same flux density as the Gaussian model (dashed line). In the figure, we bin the data in $(u,v)$ radius, by putting 4110 visibility points in one bin and display weighted averages of the visibility amplitudes for illustration purposes only; when fitting the model, all visibility points were used. Because of the large uncertainty of the visibility measurements on the longest baselines, we use the size estimate as an upper limit and thus we can derive the minimum rest frame brightness temperature of J0131$-$0321 as

\begin{equation}
T_\rmn{B}\ge1.22 \times 10^{12}\frac{S}{\vartheta^2\nu^2}(1+z) \rmn{\,K,}
\end{equation}
where $S$ is the flux density measured in Jy, $\vartheta$ is FWHM size of the component in mas, $z$ is the redshift, and $\nu$ is the observing frequency in GHz. The resulting lower limit to the brightness temperature is $T_\rmn{B}\ge2.8\times10^{11}$\,K. This brightness temperature is an order of magnitude higher than those of the other high-redshift ($z\ga4.5$), non-blazar radio-loud AGN \citep{fiveqso}, and it also exceeds that of SDSS J114657.79+403708.6 \citep{fiveqso}, a blazar \citep{1146_Ghisellini} at $z=5.0$ by two orders of magnitude.
The 1.7 GHz monochromatic radio luminosity of the source is $3\times 10^{27}$\,W\,Hz$^{-1}$. This is comparable to those of the five $4.5< z< 5$ radio-loud quasars studied by VLBI by \cite{fiveqso} and the three $z\ga5$ blazars \citep{0906_Romani,fiveqso,1026_Frey}.\footnote{In the case of Q0906+6930, we calculated the radio luminosity using data from the NRAO VLA Sky Survey \citep[NVSS, ][]{nvss} catalogue.}

Assuming an intrinsic brightness temperature for the source ($T_\rmn{B}^\rmn{int}$) and knowing a lower limit to its apparent brightness temperature, a lower limit to the Doppler factor ($\delta= T_\rmn{B}/T_\rmn{B}^\rmn{int}$) can be estimated. Often the equipartition brightness temperature \citep[$5\times10^{10}$\,K,][]{readhead} is used as the intrinsic brightness temperature of the source. This assumed value indicates moderate Doppler boosting factor as a lower limit, $\delta \ga 6$. 

The flux density of the source measured in our 1.7-GHz VLBI experiment is significantly larger than the value given in the 1.4-GHz FIRST catalogue, $33.7$\,mJy and the value given in the NVSS catalogue \citep{nvss}, $31.4$\,mJy. This can be most straightforwardly explained with variability: J0131$-$0321 was almost twice as bright at 2014.9 as it was at 2009.2 (the mean epoch of the FIRST observation) and at 1993.9 (the mean epoch of the NVSS observation). Alternatively, an extremely inverted spectrum would be needed to explain the different flux density values. The observing frequency of the FIRST survey, $1.4$\,GHz, and of our EVN observation, $1.7$\,GHz correspond to $\sim8.7$\,GHz and $\sim10.3$\,GHz in the rest frame of the source, respectively. The spectral index ($\alpha$ defined as $S\sim\nu^\alpha$) between these frequencies would be $\sim3.8$, exceeding the canonical value of $2.5$ for a single, optically thick, synchrotron self-absorbed source \citep{rybicki}. (Spectral index values lower than $2.5$ straightforwardly explained as the result of inhomogeneity in the sources.) Additionally, there is no sign of such inverted spectrum across the eight IFs of our EVN observation. The flux density values measured at the different IFs agree within the errors. Thus, the more likely scenario is that J0131$-$0321 brightened considerably with respect to its FIRST flux density. Such variability is expected from a beamed, blazar-type source.
The flux density of the phase-reference caibrator in our EVN observation, $(97.6\pm 0.3)$\,mJy, is lower than the flux density values given for the source in the NVSS and FIRST catalogues, $148.4$\,mJy and $125.7$\,mJy, respectively. Thus, it is unlikely that systematic (e.g. instrumental) amplitude calibration errors have led to the high flux density value of the target source, J0131$-$0321.

\section{Discussion}
\cite{newest} conducted {\it Swift} satellite observation of J0131$-$0321 between 2014 Oct 23 and 2014 Dec 9. They described the optical--UV--X-ray spectral energy distribution (SED) of the source with a model containing a jet, an accretion disc, and a torus \citep{Ghisellini_model}. They concluded that the viewing angle of the jet is very close to the line of sight, $\theta \sim 3^\circ-5^\circ$. Our 1.7-GHz, exploratory e-EVN observation revealed that the source is compact with relatively high brightness temperature and moderate Doppler boosting factor. Assuming there is no significant misalignment between the scales probed via SED fitting and the radio-emitting jet observed with VLBI at 1.7\,GHz, using the viewing angle and the lower limit to the Doppler factor, we can estimate the Lorentz factor of the jet ($\gamma$):
\begin{equation}
\delta=\frac{1}{\gamma(1-\beta\cos\theta)}\rmn{,}
\end{equation}
where $\beta$ is the bulk speed of the jet material given in the units of the speed of light, $c$. The resulting Lorentz factor is $\gamma\sim 3$ (and $\beta\sim0.95$). 
In the model of \cite{Ghisellini_model}, the Lorentz factor is tied to the viewing angle ($\gamma = \theta^{-1}$), and thus is more than four times larger than ours. The size of the radio emitting region ($\le 4.9$\,pc) is less than eight times the dissipation radius given by \cite{newest} ($\sim2 \times 10^{16}$\,m, corresponding to a linear size of $\sim 0.65$\,pc). Thus, to explain the existence of the two different Lorentz factors, the jet would have to decelerate significantly within this relatively small distance.

Alternatively, since our size estimate is an upper limit, and thus the Doppler factor estimate is a lower limit, using the Lorentz factor ($\gamma = 13$) and the viewing angle ($3^\circ-5^\circ$) from the model of \cite{newest}, the Doppler factor can be calculated to be in the range between $11$ and $18$. Assuming the equipartition brightness temperature for the intrinsic brightness temperature of the source, this would imply a measured brightness temperature in the range of $(5.5-9.0) \times 10^{11}$\,K. For the same flux density that we measured, this could be achieved if the radio emitting source size would be in the range between $0.25$\,mas$-0.4$\,mas corresponding to $1.5$\,pc$-2.5$\,pc in the frame of the source. (Or for the same source size, the flux density should be $\sim 2-3$ times larger, which is however in disagreement with our measurement.)

Finally, we note that the X-ray and radio observations were performed relatively close to each other but they were not strictly simultaneous, thus flux density variability of the source might hinder the comparison of the models deduced from the observations. In radio regime, the source variability is indeed detected on a longer timescale of a few years between our EVN observation and the FIRST data. Apart from the SED fitting, the blazar classification of the source is also supported by its flux density variability.

The three previously known $z\ga5$ blazars have black hole masses of $(2-5)\times 10^9M_\odot$. According to \cite{disc}, the mass of the black hole in J0131$-$0321 is $\sim 2.7\times 10^9 M_\odot$, but can be as large as $\sim 4\times 10^9M_\odot$. However, \cite{newest} assuming a \cite{shakura_sunyaev} disk model, determined an even larger black hole mass of $(1.1\pm 0.2) \times 10^{10} M_\odot$. This larger mass value no longer requires super-Eddington accretion to produce the bolometric luminosity of $L_\rmn{bol}=(1.1\pm0.2)\times10^{41}$\,W, the Eddington ratio would be $\sim 0.8$. 
The existence of supermassive black holes with few billion solar masses at early cosmological epochs \citep[e.g. at redshift $z\ga6$,][]{highmass_highredshift} poses a challenge to Eddington-limited black hole growth formation model. \cite{newest} showed that a highly spinning black hole with low accretion disk efficiency is able to explain the existence of a black hole with $10^{10} M_\odot$ at $z=5.18$. On the other hand, it could also be that the seeds of such supermassive black holes had initially been massive ($100-10^5 M_\odot$) and/or grew via super-Eddington accretion \citep{BH_form}.

\section{Summary}
We observed the newly discovered $z=5.18$ radio-bright quasar, J0131$-$0321 with EVN at $1.7$\,GHz. It was detected as a single compact radio source. We calculated a lower limit to the brightness temperature, $T_\rmn{B}\ge2.8 \times 10^{11}$\,K. Assuming the equipartition brightness temperature as the intrinsic one, the observed value indicates moderate Doppler boosting, $\delta \ga 6$. The flux density of the source in our EVN observation is almost twice as high as the the value given in the FIRST catalogue, measured $\sim5$ years earlier. Such flux density variability and the relatively high brightness temperature are indicative of the blazar nature of the source. This is in agreement with the results of \cite{newest}, who analysed the optical, UV, and X-ray measurements of the source.

VLBI observations at higher radio frequencies (e.g. $5$\,GHz) can provide a more robust limit on the source compactness and its brightness temperature. Additional constraint on the brightness temperature, thus on the Doppler factor and the blazar nature of the source, can be obtained from variability brightness temperature \citep{var_TB} provided by radio flux density monitoring of the source.

\section*{Acknowledgments}
We are grateful to the chair of the EVN Program Committee, Tom Muxlow, for granting us short exploratory e-VLBI observing time in 2014 Dec. The European VLBI Network is a joint facility of European, Chinese, South African and other radio astronomy institutes funded by their national research councils. The e-VLBI research infrastructure in Europe was supported by the European Community's Seventh Framework Programme (FP7/2007-2013) under grant agreement RI-261525 NEXPReS. The research leading to these results has received funding from the European Commission Seventh Framework Programme (FP/2007-2013) under grant agreement no. 283393 (RadioNet3). This research was supported by the Hungarian Scientific Research Fund (OTKA NN110333), and the China--Hungary Collaboration and Exchange Programme by the International Cooperation Bureau of the Chinese Academy of Sciences (CAS). TA thanks for the grant support from the China Ministry of Science and Technology under grant No. 2013CB837900.


\begin{thebibliography}{99}

\bibitem[\protect\citeauthoryear{{Beasley} \& {Conway}}{{Beasley} \&
  {Conway}}{1995}]{phase-ref}
{Beasley} A.~J.,  {Conway} J.~E.,  1995, in {Zensus} J.~A.,  {Diamond} P.~J.,
  {Napier} P.~J.,  eds, ASP Conf. Ser. Vol. 82, Very Long Baseline Interferometry and the VLBA.
  Astron. Soc. Pac., San Francisco, 
p.~327

\bibitem[\protect\citeauthoryear{{Becker}, {White} \& {Helfand}}{{Becker}
  et~al.}{1995}]{first}
{Becker} R.~H.,  {White} R.~L.,    {Helfand} D.~J.,  1995, ApJ, 450, 559

\bibitem[\protect\citeauthoryear{{Condon}, {Cotton}, {Greisen}, {Yin},
  {Perley}, {Taylor} \& {Broderick}}{{Condon} et~al.}{1998}]{nvss}
{Condon} J.~J.,  {Cotton} W.~D.,  {Greisen} E.~W.,  {Yin} Q.~F.,  {Perley}
  R.~A.,  {Taylor} G.~B.,    {Broderick} J.~J.,  1998, AJ, 115, 1693

\bibitem[\protect\citeauthoryear{{Diamond}}{{Diamond}}{1995}]{data_reduc}
{Diamond} P.~J.,  1995, in {Zensus} J.~A.,  {Diamond} P.~J.,   {Napier} P.~J.,
  eds, ASP Conf. Ser. Vol. 82, Very Long Baseline Interferometry and the VLBA.
  Astron. Soc. Pac., San Francisco,
p.~227

\bibitem[\protect\citeauthoryear{{Eisenstein}, {Weinberg}, {Agol}, {Aihara},
  {Allende Prieto}, {Anderson}, {Arns}, {Aubourg}, {Bailey}, {Balbinot} \& et
  al.}{{Eisenstein} et~al.}{2011}]{sdss}
{Eisenstein} D.~J. et al., 2011, AJ, 142, 72

\bibitem[\protect\citeauthoryear{{Fan}, {Strauss}, {Schneider}, {Becker},
  {White}, {Haiman}, {Gregg}, {Pentericci} \& et al.}{{Fan}
  et~al.}{2003}]{highmass_highredshift}
{Fan} X. et al., 2003, AJ, 125, 1649

\bibitem[\protect\citeauthoryear{{Fomalont}}{{Fomalont}}{1999}]{error}
{Fomalont} E.~B.,  1999, in {Taylor} G.~B.,  {Carilli} C.~L.,   {Perley} R.~A.,
   eds, ASP Conf. Ser. Vol. 180, Synthesis Imaging in Radio Astronomy II.
Astron. Soc. Pac. San Francisco, p.~301

\bibitem[\protect\citeauthoryear{{Frey}, {Gurvits}, {Paragi} \&
  {Gab{\'a}nyi}}{{Frey} et~al.}{2008}]{double_Frey}
{Frey} S.,  {Gurvits} L.~I.,  {Paragi} Z.,    {Gab{\'a}nyi} K.~{\'E.},  2008,
  A\&A, 484, L39

\bibitem[\protect\citeauthoryear{{Frey}, {Paragi}, {Gurvits}, {Cseh} \&
  {Gab{\'a}nyi}}{{Frey} et~al.}{2010}]{fiveqso}
{Frey} S.,  {Paragi} Z.,  {Gurvits} L.~I.,  {Cseh} D.,    {Gab{\'a}nyi}
  K.~{\'E}.,  2010, A\&A, 524, A83
  
  \bibitem[\protect\citeauthoryear{{Frey}, {Paragi}, {Fogasy} \&
  {Gurvits}}{{Frey} et~al.}{2015}]{1026_Frey}
{Frey} S.,  {Paragi} Z.,  {Fogasy} J.~O.,    {Gurvits} L.~I.,  2015, MNRAS,
  446, 2921

\bibitem[\protect\citeauthoryear{{Ghisellini} \& {Tavecchio}}{{Ghisellini} \&
  {Tavecchio}}{2009}]{Ghisellini_model}
{Ghisellini} G.,  {Tavecchio} F.,  2009, MNRAS, 397, 985

\bibitem[\protect\citeauthoryear{{Ghisellini}, {Sbarrato}, {Tagliaferri},
  {Foschini}, {Tavecchio}, {Ghirlanda}, {Braito} \& {Gehrels}}{{Ghisellini}
  et~al.}{2014}]{1146_Ghisellini}
{Ghisellini} G.,  {Sbarrato} T.,  {Tagliaferri} G.,  {Foschini} L.,
  {Tavecchio} F.,  {Ghirlanda} G.,  {Braito} V.,    {Gehrels} N.,  2014, MNRAS,
  440, L111

\bibitem[\protect\citeauthoryear{{Ghisellini}, {Tagliaferri}, {Sbarrato} \&
  {Gehrels}}{{Ghisellini} et~al.}{2015}]{newest}
{Ghisellini} G.,  {Tagliaferri} G.,  {Sbarrato} T.,    {Gehrels} N.,  2015, MNRAS, in press (arXiv:1501.07269)

\bibitem[\protect\citeauthoryear{{Greisen}}{{Greisen}}{2003}]{aips}
{Greisen} E.~W.,  2003,
Information Handling in Astronomy - Historical Vistas,
  285, 109

\bibitem[\protect\citeauthoryear{{Gurvits}}{{Gurvits}}{1999}]{gurvits}
{Gurvits} L.~I.,  1999, in {van Harleem} M.~P.,  ed., Perspectives on Radio
  Astronomy: Science with Large Antenna Arrays. ASTRON, Dwingleoo, p.~183

\bibitem[\protect\citeauthoryear{{H\"ogbom}}{{H\"ogbom}}{1979}]{clean}
{H\"ogbom} J.~A.,  1979, A\&AS, 36, 173

\bibitem[\protect\citeauthoryear{{Hovatta}, {Valtaoja}, {Tornikoski} \&
  {L{\"a}hteenm{\"a}ki}}{{Hovatta} et~al.}{2009}]{var_TB}
{Hovatta} T.,  {Valtaoja} E.,  {Tornikoski} M.,    {L{\"a}hteenm{\"a}ki} A.,
  2009, A\&A, 494, 527

\bibitem[\protect\citeauthoryear{{Keimpema}, {Kettenis}, {Pogrebenko},
  {Campbell}, {Cim{\'o}}, {Duev}, {Eldering}, {Kruithof}, {van Langevelde},
  {Marchal}, {Molera Calv{\'e}s}, {Ozdemir}, {Paragi}, {Pidopryhora}, {Szomoru}
  \& {Yang}}{{Keimpema} et~al.}{2015}]{soft_corr}
{Keimpema} A. et al., 2015, Exp.
  Astron., in press (arXiv:1502.00467)

\bibitem[\protect\citeauthoryear{{Kovalev}, {Kellermann}, {Lister}, {Homan},
  {Vermeulen}, {Cohen}, {Ros}, {Kadler}, {Lobanov}, {Zensus}, {Kardashev},
  {Gurvits}, {Aller} \& {Aller}}{{Kovalev} et~al.}{2005}]{kovalev}
{Kovalev} Y.~Y. et al.,  2005, AJ, 130, 2473

\bibitem[\protect\citeauthoryear{{McLure} \& {Dunlop}}{{McLure} \&
  {Dunlop}}{2004}]{mclure}
{McLure} R.~J.,  {Dunlop} J.~S.,  2004, MNRAS, 352, 1390

\bibitem[\protect\citeauthoryear{{Momjian}, {Carilli} \& {McGreer}}{{Momjian}
  et~al.}{2008}]{double_Momjian}
{Momjian} E.,  {Carilli} C.~L.,    {McGreer} I.~D.,  2008, AJ, 136, 344

\bibitem[\protect\citeauthoryear{{Readhead}}{{Readhead}}{1994}]{readhead}
{Readhead} A.~C.~S.,  1994, ApJ, 426, 51

\bibitem[\protect\citeauthoryear{{Romani}}{{Romani}}{2006}]{0906_Romani}
{Romani} R.~W.,  2006, AJ, 132, 1959

\bibitem[\protect\citeauthoryear{{Rybicki} \& {Lightman}}{{Rybicki} \&
  {Lightman}}{1986}]{rybicki}
{Rybicki} G.~B.,  {Lightman} A.~P.,  1986, {Radiative Processes in
  Astrophysics}.
Wiley-VCH, New York

\bibitem[\protect\citeauthoryear{{Sbarrato}, {Tagliaferri}, {Ghisellini},
  {Perri}, {Puccetti}, {Balokovi{\'c}}, {Nardini}, {Stern} \& et
  al.}{{Sbarrato} et~al.}{2013a}]{1026_Sbarrato}
{Sbarrato} T. et al., 2013a, ApJ,
  777, 147

\bibitem[\protect\citeauthoryear{{Sbarrato}, {Ghisellini}, {Nardini},
  {Tagliaferri}, {Greiner}, {Rau} \& {Schady}}{{Sbarrato}
  et~al.}{2013b}]{cand_Sbarrato}
{Sbarrato} T.,  {Ghisellini} G.,  {Nardini} M.,  {Tagliaferri} G.,  {Greiner}
  J.,  {Rau} A.,    {Schady} P.,  2013b, MNRAS, 433, 2182

\bibitem[\protect\citeauthoryear{{Shakura} \& {Sunyaev}}{{Shakura} \&
  {Sunyaev}}{1973}]{shakura_sunyaev}
{Shakura} N.~I.,  {Sunyaev} R.~A.,  1973, A\&A, 24, 337

\bibitem[\protect\citeauthoryear{{Shepherd}, {Pearson} \& {Taylor}}{{Shepherd}
  et~al.}{1994}]{difmap} {Shepherd} M.~C., {Pearson} T.~J., {Taylor} 
G.~B., 1994, BAAS, 26, 987

\bibitem[\protect\citeauthoryear{{Szomoru}}{{Szomoru}}{2008}]{eEVN}
{Szomoru} A.,  2008, in Proceedings of Science, PoS(IX EVN Symposium)040

\bibitem[\protect\citeauthoryear{{Trakhtenbrot} \& {Netzer}}{{Trakhtenbrot} \&
  {Netzer}}{2012}]{Trak}
{Trakhtenbrot} B.,  {Netzer} H.,  2012, MNRAS, 427, 3081

\bibitem[\protect\citeauthoryear{{Volonteri}}{{Volonteri}}{2012}]{volonteri}
{Volonteri} M.,  2012, Science, 337, 544

\bibitem[\protect\citeauthoryear{{Wright}, {Eisenhardt}, {Mainzer}, {Ressler},
  {Cutri}, {Jarrett}, {Kirkpatrick}, {Padgett} \& et al.}{{Wright}
  et~al.}{2010}]{wise}
{Wright} E.~L. et al., 2010, AJ, 140, 1868

\bibitem[\protect\citeauthoryear{{Wu}, {Hao}, {Jia}, {Zhang} \& {Peng}}{{Wu}
  et~al.}{2012}]{sdss_selection}
{Wu} X.-B.,  {Hao} G.,  {Jia} Z.,  {Zhang} Y.,    {Peng} N.,  2012, AJ, 144, 49

\bibitem[\protect\citeauthoryear{{Wyithe} \& {Loeb}}{{Wyithe} \&
  {Loeb}}{2012}]{BH_form}
{Wyithe} J.~S.~B.,  {Loeb} A.,  2012, MNRAS, 425, 2892

\bibitem[\protect\citeauthoryear{{Yi}, {Wang}, {Wu}, {Yang}, {Bai}, {Fan},
  {Brandt}, {Ho}, {Zuo}, {Kim}, {Wang}, {Yang}, {Zhang}, {Wang}, {Wang}, {Ai},
  {Fan}, {Chang}, {Wang}, {Lun} \& {Xin}}{{Yi} et~al.}{2014}]{disc}
{Yi} W.-M. et al.,  2014, ApJ, 795, L29

\bibitem[\protect\citeauthoryear{{Zhang}, {Wang}, {Bai}, {Zhang}, {Wang},
  {Liu}, {Zhao} \& {Chen}}{{Zhang} et~al.}{2014}]{FOSC}
{Zhang} J.-J.,  {Wang} X.-F.,  {Bai} J.-M.,  {Zhang} T.-M.,  {Wang} B.,  {Liu}
  Z.-W.,  {Zhao} X.-L.,    {Chen} J.-C.,  2014, AJ, 148, 1

\end{thebibliography}

\bsp

\label{lastpage}

\end{document}